\input{aipcheck}

\documentclass[
    ,final            
  ]
  {aipproc}

\layoutstyle{8x11double}

\begin{document}

\title{RR Lyrae Pulsation Theory}


\author{Marcella Marconi}{address={INAF Osservatorio Astronomico di Capodimonte, Vicolo 
Moiariello 16, 80131 Napoli, Italy}
}

\begin{abstract}
RR Lyrae stars play an important role as distance indicators and stellar
population tracers. In this context the construction of accurate pulsation
models is crucial to understand the observed properties and to
constrain the intrinsic stellar parameters of these pulsators.
The physical mechanism driving pulsation in RR Lyrae stars has been known since the
middle of the 20th century and many efforts have been performed during the
last few decades in the construction of more and more refined pulsation
models.
In particular, nonlinear pulsation models including a nonlocal time-dependent
treatment of convection, such as the ones originally developed in Los Alamos
in the seventies, allow us to reproduce all the relevant observables of
stellar radial pulsation and to establish accurate relations and methods to 
contrain
the intrinsic stellar properties and the distances of these variables.
The most recent results on RR Lyrae pulsation obtained through these kinds of
models will be presented and a few still debated problems will be discussed.

\end{abstract}

\maketitle


\section{Introduction}

RR Lyrae are low mass central He burning stars belonging to the 
Horizontal Branch (HB) evolutionary phase in the Hertzprung-Russell (HR) 
diagram. They are important tracers of the oldest stellar populations and the 
knowledge of their properties can provide relevant constraints to several 
important aspects of stellar evolution and cosmology.
RR Lyrae are also important distance indicators for Pop. II systems through
the absolute visual magnitude versus metallicity ($M_V(RR)-[Fe/H]$) relation
but also thanks to the existence of a Period-Luminosity (PL) correlation in the Near Infrared 
bands. On this basis, the theoretical predictions through accurate pulsation 
models of the observed properties of RR Lyrae stars are of crucial importance for our understanding
of many relevant astronomical issues.
RR Lyrae are characterized by a variety of observed behaviours. As well known, with a period ranging from $\sim$ 0.3 to 1.0 d and a pulsation amplitude in the V band always smaller than 2 mag, they are divided into two main subgroups: 
RRab pulsating in the fundamental mode and RRc pulsating in the first overtone 
mode. The first class shows asymmetric light curves with the pulsation amplitude decreasing with
 the period, while the latter class is characterized by shorter periods, smaller 
amplitudes and more symmetric light curves than the RRab. In the 
period-amplitude diagram (Bailey diagram) RRab 
stars have a linear behaviour, whereas first overtone pulsators form a 
bell-shaped sequence. 
They are located on the HB, with an absolute magnitude in the V banding ranging from $\sim$ 0 to $\sim$ 1 mag, and their distribution in color in a given globular cluster reflects the distribution in mass resulting from the mass loss phenomenon in the previous red giant branch (RGB) evolutionary phase. RR Lyrae in Galactic Globular Clusters (GGCs) show a phenomenon known 
as the Oosterhoff Dichotomy dating back to Oosterhoff (1939). According to this phenomenon GGCs 
are separated into two groups: the Oosterhoff I (OoI) with average period of RRab 
($<P_{ab}>$) close to 0.55 d, a larger number of RRab than RRc and intermediate
 metallicity; the Oosterhoff II (OoII) with $<P_{ab}>$ around 0.65 d, 
similar number of RRab and RRc and low metallicity.
Accurate pulsation models should be able to reproduce all these properties as well 
as the absolute visual magnitude versus metallicity relation and the PL correlation in the NIR 
bands, but also additional observational features such as the behaviour of 
mixed mode pulsators and the Blazhko effect (see the review by G. Kovacs, this volume).
In this paper we present a review of RR Lyrae pulsation models and summarize some 
relevant results obtained from nonlinear convective radial hydrocodes.
In Section 2 we briefly review the physical mechanism driving pulsation in 
RR Lyrae stars; in Section 3 we summarize the theoretical efforts developed by several authors to intepret the RR Lyrae pulsation properties 
from a linear 
non adiabatic approach to nonlinear convective models; in Section 4 we discuss some significant results concerning RR Lyrae  with 
implications for stellar astrophysics and the distance scale calibration;  
 in Section 5 a number of still open questions are mentioned and some future 
perspectives are outlined. The conclusions close the paper.

\section{RR Lyrae pulsation mechanisms}

Since the pioneering investigations by  Eddington (1926) and  Zhevakin (1959) it was clear that the physical phenomenon driving pulsation in
variable stars lying in the classical pulsation instability strip, including RR Lyrae 
stars, was a valve mechanism mainly efficient in the second ionization region 
of He (30,000-60,000 K). This {\it k mechanism} was later confirmed by Baker 
\& Kippenhahn (1962) and Cox (1963), whereas Christy (1962) first 
demonstrated that the HeI and H ionization layers also play a role in driving 
pulsation, as subsequently confirmed by Bono \& Stellingwerf (1994).
The need for a non-adiabatic theory to model the {\it $\kappa$ mechanism} and to treat
 the problem of stellar stability is evident. Indeed from a linear adiabatic approach 
only the pulsation periodicity can be predicted.

\section{From linear non-adiabatic to nonlinear convective models}

The development of accurate pulsation models for RR Lyrae stars has known a tremendous improvement duriong the last few decades, with the inclusion of more and more sophisticated physical and numerical ingredients and their continuous update.

\subsection{Linear non-adiabatic models}

Several authors computed linear non-adiabatic radial models of RR Lyrae stars, 
both in a radiative regime (e.g. Cox 1963; Castor 1971; van Albada \& 
Baker 1971; Iben \& Huchra 1971; Glasner \& Buchler 1993; D\'ek\`any et al. 
2008) and including convection (e.g. Tuggle \& Iben 1973; Baker \& Gough 
1979; Xiong 1982; Xiong et al. 1998). These models were able to predict the
 topology of the RR Lyrae instability strip. In particular radiative linear 
non-adiabatic models provided only the location  of the blue edge (see e.g. 
Iben \& Huchra 1971), whereas the convective ones were able to predict also an 
approximate evaluation of the effective temperature of the red edge (Tuggle \& 
Iben 1973; Xiong et al. 1998) because it is the quenching effect of convection on 
pulsation that determines the occurrence of the red edge.
 A very important result of linear non 
adiabatic convective models was obtained by van Albada \& Baker (1971), who 
predicted linear relations (logarithmic scale) between the pulsation period 
and the stellar mass, luminosity and effective temperature for both the 
fundamental and the first overtone mode. These relations establish an 
important link between pulsational and evolutionary parameters and 
represented the milestone of any study aimed at investigating the evolutionary 
behaviour of RR Lyrae through the analysis of their pulsation characteristics.
In a subsequent paper (van Albada \& Baker 1973) they predicted the 
efficiency 
of an {\it hysteresis mechanism} in determining the pulsation mode in the OR 
region: RR Lyrae that enter the instability strip from the blue keep pulsating in the first overtone mode, while the ones that come from the red continue to pulsate in the fundamental mode. This idea implies that the morphology of the evolutionary tracks determines the pulsation mode in the OR region, and was at the basis of several subsequent papers aiming at 
explaining the Oosterhoff dichotomy (see e.g. Bono et al. 1995 and references therein).
A time-dependent generalization of the mixing-length theory was adopted by 
Baker \& Gough (1979) in the construction RR Lyrae nonadiabatic models. 
These authors 
studied the effect of 
including the interaction between pulsation and convection on their stability analysis.
A similar approach 
was followed more recently by Xiong et al. (1998) on the basis of 
a non-local and time-dependent treatment of convection.

\subsection{Nonlinear models}

When the small oscillation approximation is released and the fundamental 
equations of the pulsating stellar enevelope are not linearized, we are in the 
position to predict the full amplitude behaviour of models at their limit cycle.In particular nonlinear models are able to provide information not only on the pulsation period and the instability strip boundaries, but also on the pulsation amplitudes and on the morphological characteristics of the variation along a pulsation cycle of luminosity, radius, radial velocity, effective temperature, surface gravity.
This is true also for radiative models even if, without including convection, no 
information on the red boundary of the instability strip can be obtained and 
there are also difficulties in matching the observed light curve amplitudes 
and morphologies (Kovacs \& Kanbur 1998, Feuchtinger 1999a).
In order to properly reproduce all the observables of radial pulsation 
nonlinear convective models are required.
Since the early 80's several authors included convection in their nonlinear 
hydrocodes (Stellingwerf 1982, 1984; Gehmeyr 1992, 1993; Bono \& 
Stellingwerf 1994; Feuchtinger 1999a; Szab\`o, Kollath \& Buchler 2004).
Stellingwerf was the first to introduce a nonlocal time-dependent treatment of 
convection in RR Lyrae nonlinear pulsation models, based on the treatment of 
the transport equation by Castor (1968). Several subsequent investigations of 
RR Lyrae properties, that provided relevant results for the use of these stars as distance indicators and stellar population tracers, were based on refinements and updates of Stellingwerf's 
original code (Bono \& Stellingwerf 1994; Bono et al. 1997, 2003; Marconi et al. 2003; Di Criscienzo et al. 2004; Marconi \& 
Clementini 2005; Marconi \& Degl'Innocenti 2007; Marconi et al. 2009 in preparation).
On the other hand the hydrocode developed by Gehmeyr (1992) was based on 
an adaptive grid approach and on a nonlinear convective treatment  and was able to determine precisely the red edge position; Feuchtinger (1999a) used 
the Vienna nonlinear convective code (Feuchtinger 1999b) to predict accurate 
theoretical light and radial velocity curves, as well as to reproduce the Fourier parameters; finally 
Szab\`o et al. (2004) performed a consistent modeling of pulsation and 
evolutionary properties through an amplitude equation formalism and were able 
to reproduce the observed slope of the fundamental blue edge, as well as the 
behaviour of double mode RR Lyrae.
On this basis, we can conclude that nonlinear convective radial models are 
necessary to reproduce all the relevant pulsation observables, apart from a number 
of additional phenomena that require the inclusion of nonradial pulsation 
modes and/or the introduction of new physics. These are for example the 
Blazhko effect (see the review by G. Kovacs, this volume), the nonradial pulsation of RR 
Lyrae stars (see e.g. the contribution by A. Cox, this volume), the role of magnetic fields (see e.g. the contribution by K. Kolenberg, this volume), as well as the formation and propagation of shock waves 
(see e.g. the contributions by Chadid and Paparo, this volume). 

\section{Significant results concerning RR Lyrae}
Many important results have been obtained during the last few decades from nonlinear convective models of RR Lyrae stars, with relevant implications for both the distance scale problem and for the study of stellar populations.
In this section we will concentrate on three important topics, namely the 
topology of the RR Lyrae instability strip and of the ensuing Bailey diagram, 
the pulsation stellar mass determination from the study of RR Lyrae properties, 
and the calibration of the RR Lyrae distance scale.

\subsection{Topology of the RR Lyrae instability strip} 
Nonlinear convective pulsation models allow us to predict the complete 
topology of the instability strip for both fundamental and first overtone 
modes. Fig. 1 reports the location of the first overtone blue edge (FOBE) 
the fundamental blue edge (FBE), the first overtone red edge (FORE) and 
the fundamental red edge (FRE) in the HR diagram, as obtained on the basis of 
our nonlinear convective pulsation models (Bono, Caputo, Marconi 1995; 
but see also Bono et al. 1997) for Z=0.0001 and Y=0.24.

\begin{figure}
  \includegraphics[height=.4\textheight]{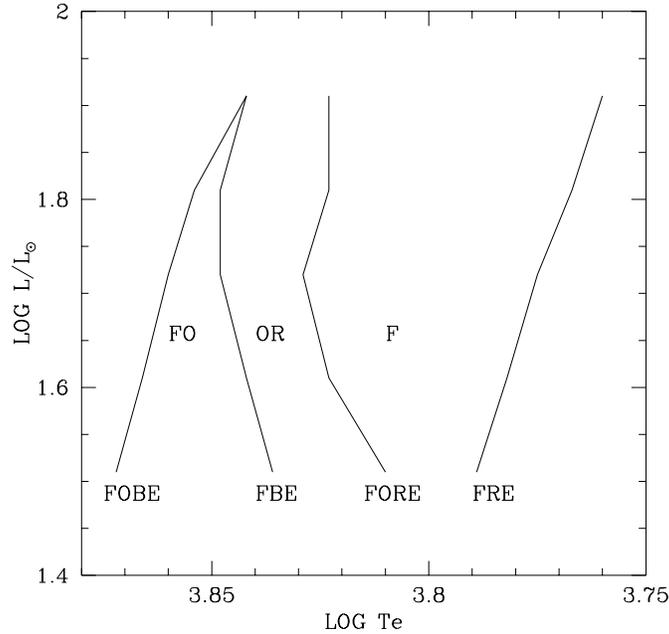}
  \caption{Topology of the fundamental and first overtone instability strip for RR Lyrae models
with Z=0.0001, Y=0.24 and M=0.65 $M_{\odot}$.}
\end{figure}

The intersection region between the fundamental and the first overtone 
instability strips is called OR region and the different population of this 
region by fundamental and first overtone pulsators, as triggered by the 
already mentioned {\it hysteresis mechanism}  is at the basis of the 
explanation of the Oosterhoff dichotomy suggested by current pulsation models 
(see Bono et al. 1995, 1997 for details). According to this 
explanation, the OR 
region is populated by $RR_{ab}$ in OoI GCs and by $RR_{c}$ in OoII GCs, so 
that the average period of $RR_{ab}$ is shorter in OoI than in OoII.
This different population of the OR region is also evident from inspection of 
Fig. 2, showing the comparison in the Bailey diagram of RR Lyrae belonging to 
the two prototype GGCs of the OoI and 
OoII classes, namely M3 and M15. The middle part of this plot, roughly 
corresponding to the OR region, is populated by high 
amplitude $RR_{ab}$ in M3 and by the decreasing branch of the $RR_{c}$ 
{\it bell} in M15.

\begin{figure}
  \includegraphics[height=.3\textheight]{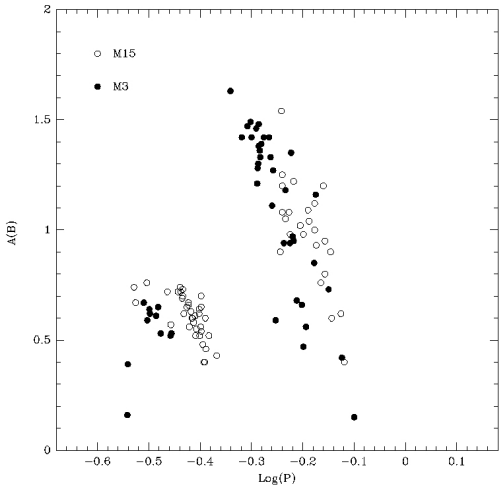}
  \caption{Comparison in the Bailey diagram between the RR Lyrae belonging to 
the two prototype GGCs of the OoI and 
OoII classes, namely M3 (filled circles) and M15 (open circles).}
\end{figure}

In other terms M15 lacks high amplitude F pulsators and M3 lacks the 
decreasing branch of the FO {\it bell} in the OR region.
The theoretical Bailey diagram based on our nonlinear convective models nicely 
reproduces the observed linear behaviour of fundamental pulsators, providing 
the basis for the derivation of Period-Luminosity-Amplitude relations (see Di Criscienzo, Marconi, Caputo 2004 and 
Bono, Caputo \& Di Criscienzo 2007 and references therein) that can in principle be used as an additional theoretical tool to infer RR Lyrae distances (see below). This is shown in Fig. 3 for a sample of GGCs with a signifacnt number of RR Lyrae (see Di Criscienzo et al. 2004 for details).

\begin{figure}
  \includegraphics[height=.5\textheight]{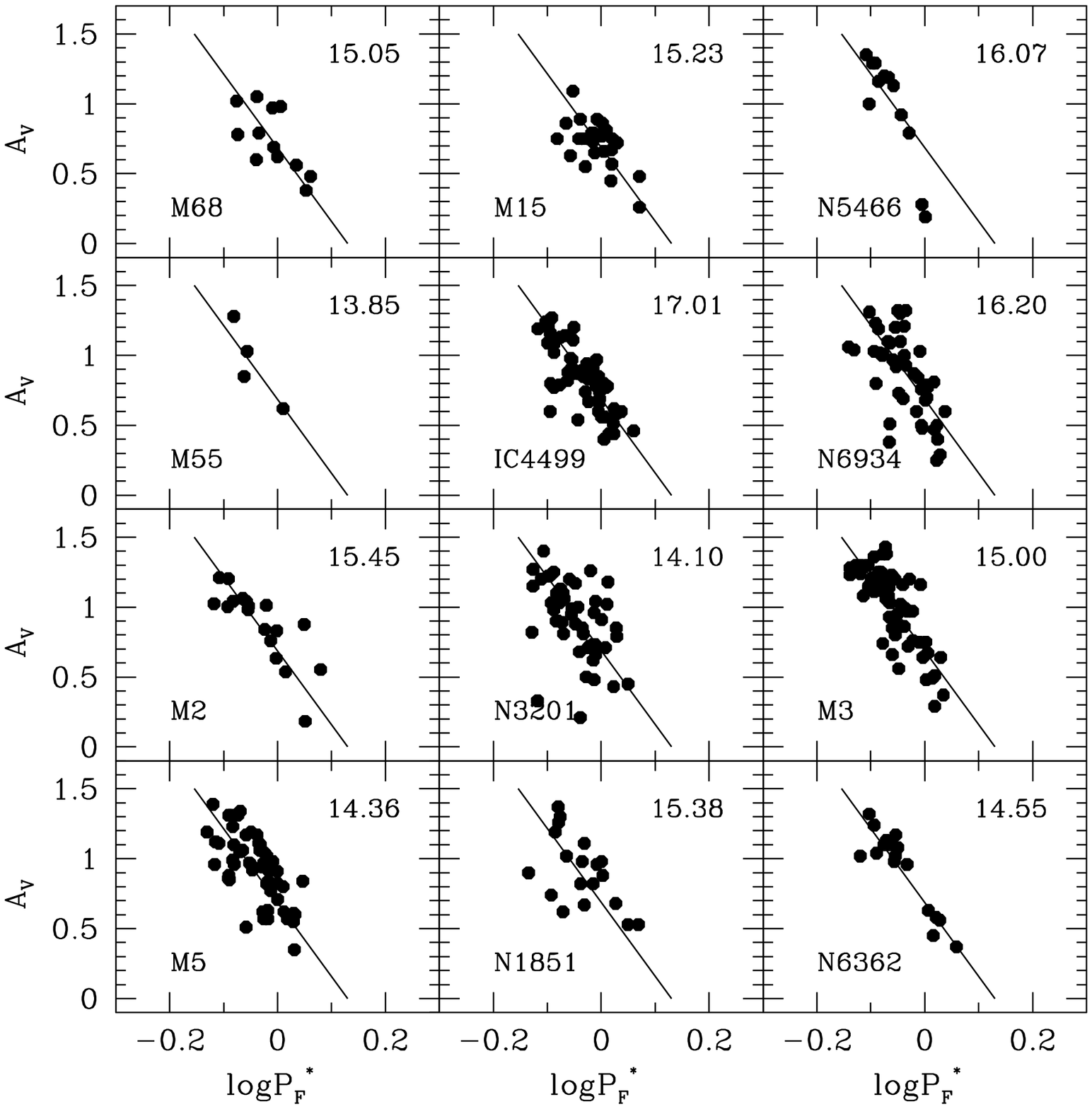}
  \caption{Application of a Period-Luminosity-Amplitude relation for fundamental pulsators at l/Hp = 1.5 to the RR Lyrae in a sample of GGCs. The solid lines are the predicted relations at constant mass. The derived apparent distance moduli are given in each panel.}
\end{figure}

\subsection{Stellar pulsation mass determination}

A {\it pulsational} mass estimate, that is independent of stellar evolution 
predictions, can be obtained either using double mode 
pulsators or through the model fitting of observed light (and radial velocity) 
curves.
The first method is based on the use of the Petersen diagram (see Petersen 
1978 for the first application to RR Lyrae stars), reporting the ratio 
between the first overtone and the fundamental period as a function of the 
fundamental period (in logarithmic scale). Many fundamental papers were based 
on the comparison between linear non-adiabatic model predictions and 
observations in the Petersen diagram, as a tool to constrain the stellar mass 
of field and cluster RR Lyrae (see e.g. Cox, Hodson \& King 1980; Cox, Hodson \& Clancy 1983). At the time these {\it pulsational} masses were found to be in good agreement with stellar evolution estimates.
However, with the update of the input physics in evolutionary computations, a 
{\it mass discrepancy} problem appeared with {\it pulsational} masses resulting to be 
systematically smaller than the evolutionary ones. This discrepancy was solved when the input physics of pulsation models 
was 
also updated, in  particular with the adoption of Livermore opacity tables 
(see Cox 1991 and Bono et al. 1996 for details). Moreover Bono et 
al. (1996) found that using nonlinear convective models the 
comparison with observations in the Petersen diagram can be used not only to 
derive the 
stellar mass but also to constrain the luminosity level.

The second method to infer a {\it pulsational} estimate of RR Lyrae mass 
 is the model fitting of observed light and, when available, radial velocity 
curves. The fact that nonlinear 
convective models allow us to predict accurate variations of all the relevant 
quantities along a pulsation cycle implies that the comparison between 
theoretical and observed variations represents a powerful tool to constrain 
the intrinsic stellar parameters including the mass. The model fitting of 
observed light curves was applied for the first time to a Cepheid in the LMC 
by Wood, Arnold \& Sebo (1997), whereas the first application to a 
RR Lyrae pulsator, the Galactic field variable U Comae, was presented by Bono, 
Castellani \& Marconi (2000). As shown in Fig. 4, in the case of U Comae, the obtained best fit model was able 
to accurately 
reproduce the light curves in the UBV filters, and also the poor sampled K 
band curve, for stellar parameters, including the mass (and the distance), in 
good agreement with the literature and the evolutionary prescriptions. 

\begin{figure}
  \includegraphics[height=.2\textheight]{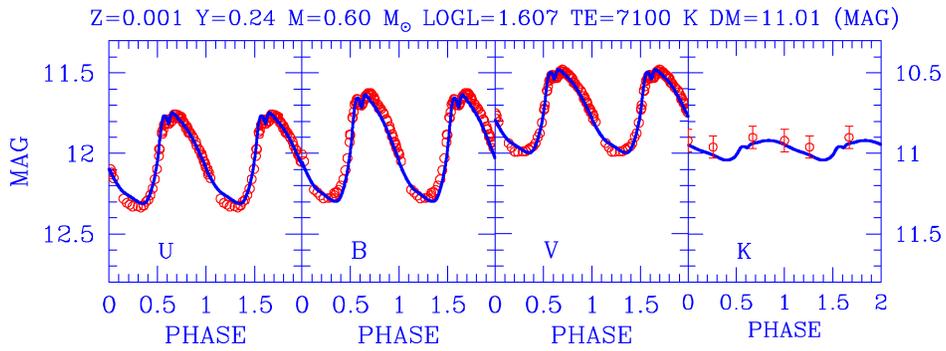}
  \caption{Model fitting of U Comae multi-band ligth curves (see Bono et al. 2000 for details.)}
\end{figure}

\begin{figure}
\includegraphics[height=.5\textheight]{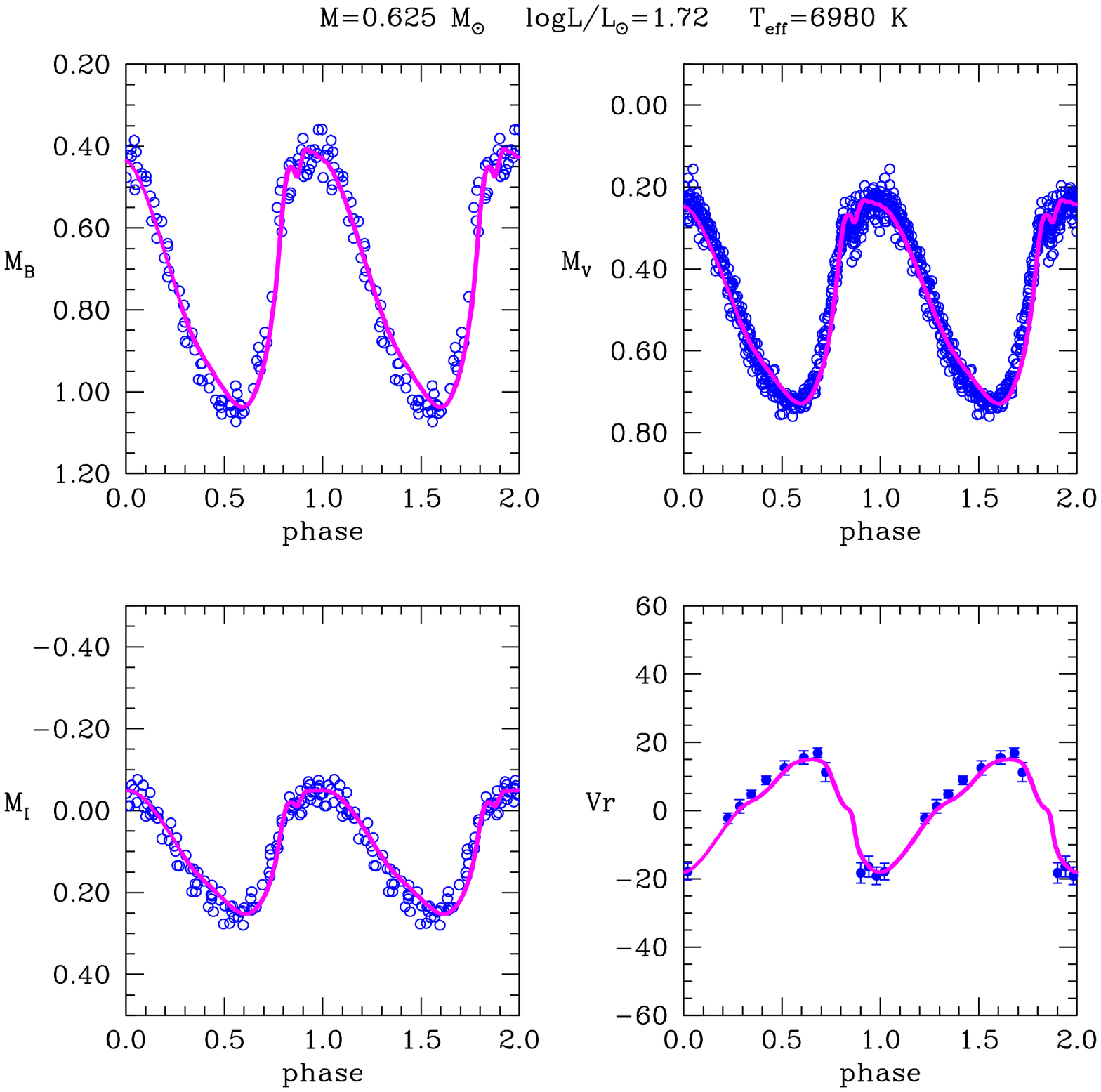}
\caption{Model fitting of light and radial velocity 
curves for the field RR Lyrae CM Leo (see Di Fabrizio et al. 2002 for details.)}
\end{figure}

A second 
example showing the successful simultaneous fit of light and radial velocity 
curves was the case of CM Leo (see Di Fabrizio et al. 2002). The comparisons between the data and the best fit model are shown in Fig. 5. 
Again the best fit model was found to have stellar parameters consistent with 
evolutionary predictions.
More recently the method has been applied also to cluster RR Lyrae. In 
Marconi \& Degl'Innocenti (2007) we presented the first application to a 
GGC performing the model fitting of the light curves of a sample of RR Lyrae 
in M3. The stellar masses of the selected pulsators were constrained and found 
to be in agreement with the evolutionary predictions. At the same time the 
resulting distance modulus is in very good agreement with independent estimates
 in the literature.
The application of the model fitting technique to more peculiar GGCs, namely 
NGC2419 and  NGC6441, is in progress (Di Criscienzo et al. 2009 in preparation;
 Marconi \& Clementini 2009 in preparation).

\subsection{Calibration of the RR Lyrae distance scale}

As already noted, RRLyrae are the most important primary distance indicators 
for Pop.II systems. Several results have been obtained from the thoeretical 
point of view through the calibration of different methods.
First of all, the model fitting technique discussed above is a powerful tool to 
infer individual RR Lyrae distances through the comparison between the apparent 
magnitudes and the best fit model intrinsic luminosity.
The results of the application of this method to a sample of RR Lyrae in  the 
LMC are shown in Fig. 6, for the fundamental and first overtone 
pulsators respectively (see Marconi \& Clementini 2005).The resulting 
distance modulus is $18.54\pm0.02$ mag and is in excellent agreement with the 
values 
obtained through the model fitting of other classes of pulsating stars in the 
LMC, namely two samples of Classical Cepheids ($18.53\pm0.05$ mag Bono, 
Castellani, Marconi 2002; $18.54\pm0.02$ mag Keller \& Wood 2002, 2006)
 and a $\delta$ Scuti pulsator ($18.48\pm0.15$ mag McNamara, Cementini \& 
Marconi 2007).

\begin{figure}
  \resizebox{20pc}{!}{\includegraphics{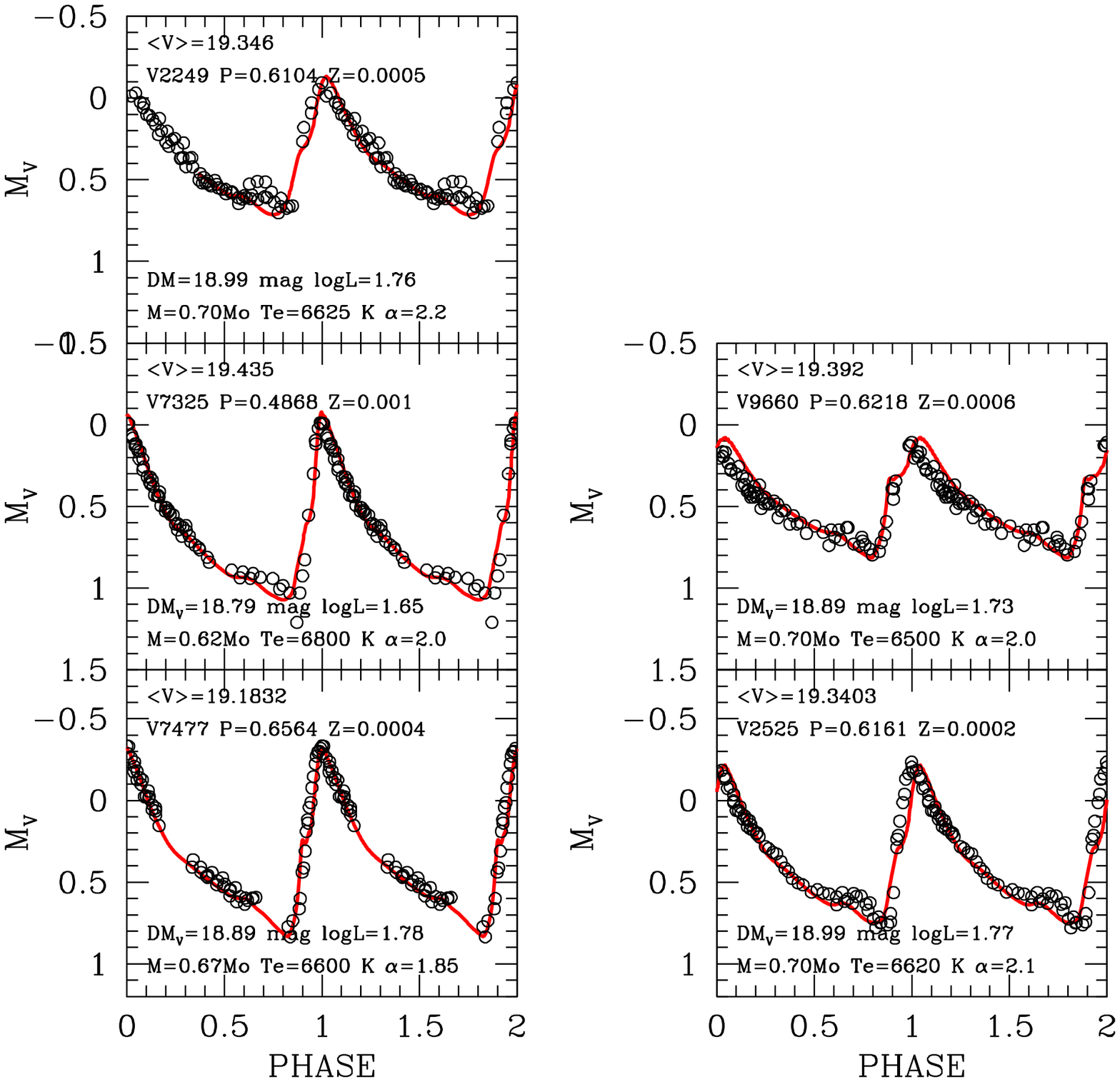}}  
  \resizebox{20pc}{!}{\includegraphics{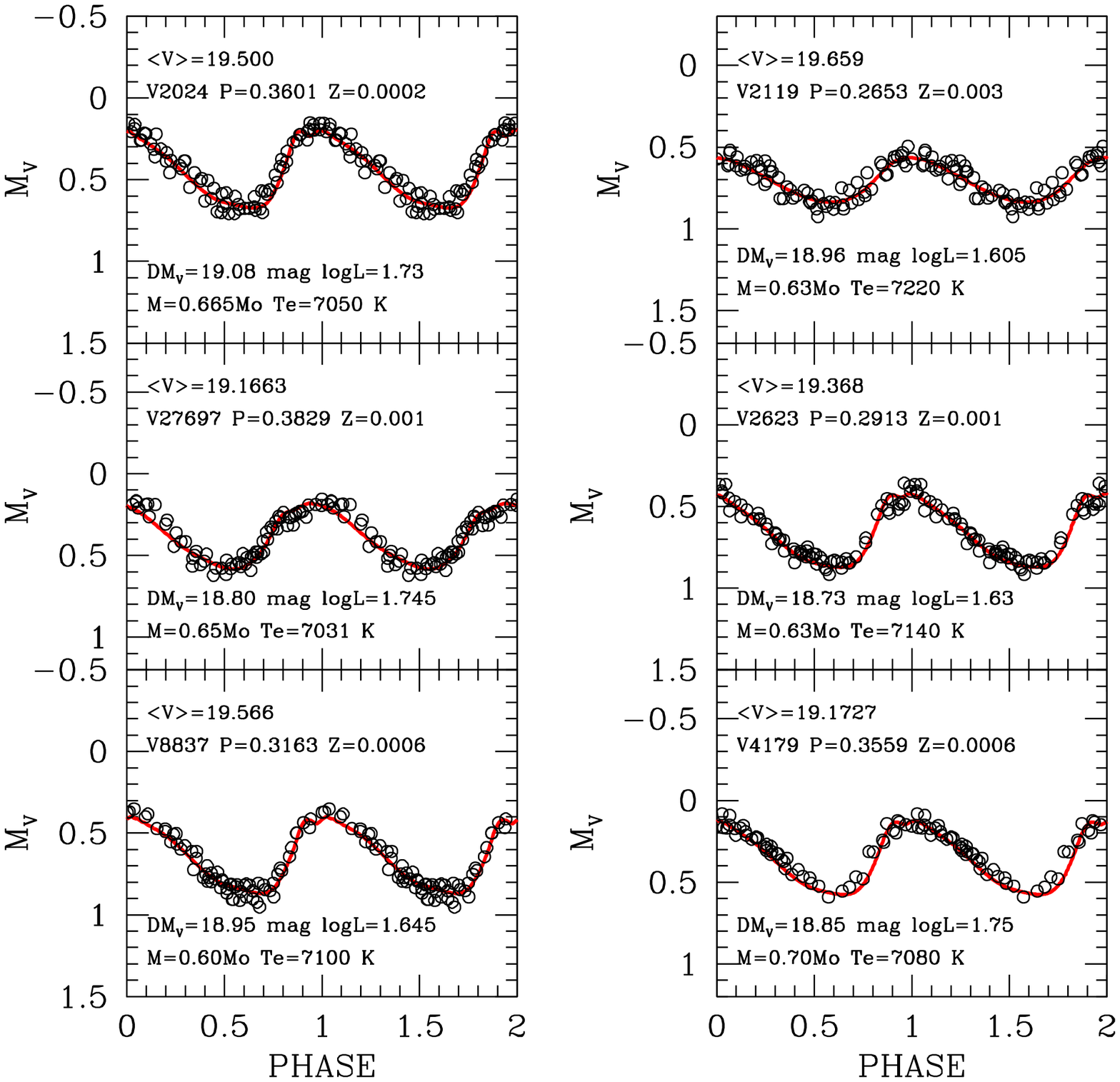}}
\caption{Model fitting of the V band light curves for a sample of 5 RRab (left panels) and 6 RRc (right panels) in the LMC.}
\end{figure}
 
A method to infer cluster RR Lyrae distances is based on the comparison 
between theory and observations in the $M_V$ versus $\log {P}$ plane (Caputo 
1997; Caputo et al. 2000; Di Criscienzo et al. 2004).
By matching the observed blue edge\footnote{The red edge is less reliable due 
to its well known sensitivity to the efficiency of convection and to the 
uncertianties still affecting the treatment of turbulent convection.} of the 
instability strip with the theoretical one, as based on an extensive set of 
nonlinear convective pulsation models, we obtain a direct estimate of the 
apparent distance modulus for each investigated cluster and, by subtracting this value
to the apparent mean 
magnitude level of the RR Lyrae, also an estimate of the absolute RR Lyrae 
magnitude (and of the true distance modulus for each investigated GGC, see Di 
Criscienzo et al. 2004 for details).
As a result of the application of this technique to a sample of GGCs containing RR Lyrae (stistically significant samples) we obtain a $M_V(RR)-[Fe/H]$ 
relation that is not linear over the whole observed metallicity range but with 
a slope that gets steeper as the metallicity increases and evidence of a 
change of the slope around [Fe/H]=-1.5 as shown in Fig. 7 (see also Caputo et al. 2000).

\begin{figure}
\includegraphics[height=.5\textheight]{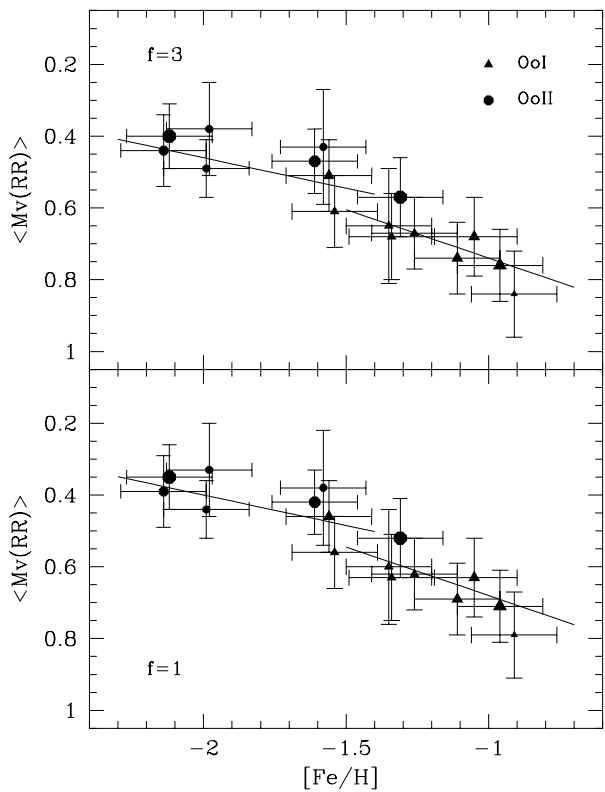}
\caption{$M_V(RR)-[Fe/H]$ relations for two assumptions on the $\alpha$ element enhancement, as obtained from application
of the FOBE fitting technique to a sample of GGCs containing rich samples of RR Lyrae (see Di Criscienzo et al. 2004 for details.)}
\end{figure}

The individual distance moduli obtained from the fitting of the blue edge in 
the $M_V$ versus $\log {P}$ plane are in excellent agreement with the values 
inferred from application of the theoretical period-Wesenheit relations in the 
BV bands based on the same set of pulsation models (see Di Criscienzo et al. 
2004 for details).
Finally, current pulsation models are able to reproduce the already quoted 
correlation between the period and the absolute magnitude in the near-infrared bands and in particular in the K (2.2 $\mu$m) band
(Bono et al. 2001, 2002, 2003), providing useful constraints on 
the PL(K)-based RR Lyrae distance scale. These theoretical predictions show 
that the metallicity affects the PL(K) relation: RR Lyrae models are found to 
obey to 
a $\log P - M_K - [Fe/H]$ relation rather than to a $\log P - M_K$ relation.
The application of these model results to the data for the prototype RR Lyr 
gives a 
distance value in excellent agreement with the HST astrometric determination, as shown in Fig. 8
(Bono et al. 2002, 2003), whereas the application to a sample of well 
studied field RR Lyrae confirms the existence of a strict correlation between 
the period, the K band magnitude, and the metallicity, as shown in Fig. 9. Moreover, correcting 
the 
apparent V magnitudes with the distances obtained from the K band theoretical 
analysis, we obtain a  $M_V(RR)-[Fe/H]$ relation consistent with the one obtained from 
the fitting of the blue edge in $M_V$ versus $\log {P}$ plane (see Bono et al. 
2003 for details).

\begin{figure}
  \includegraphics[height=.5\textheight]{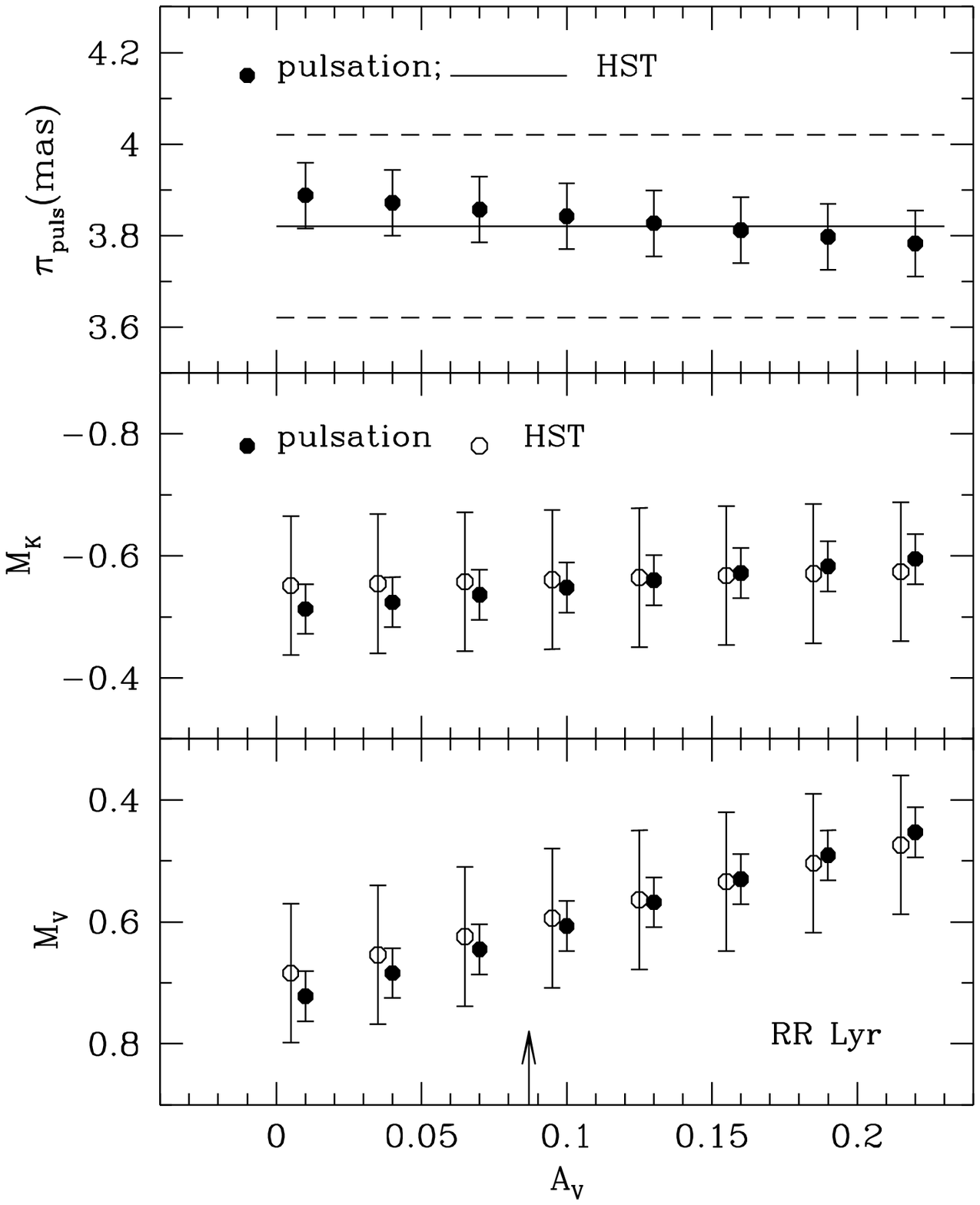}
  \caption{Comparison between the theoretical pulsation parallax based on the calibration of the K band PL and the astrometric HST result (upper panel). Similar comparisons for the K and V absolute magnitudes are also shown (middle and lower panels).}
\end{figure}

\begin{figure}
  \includegraphics[height=.4\textheight]{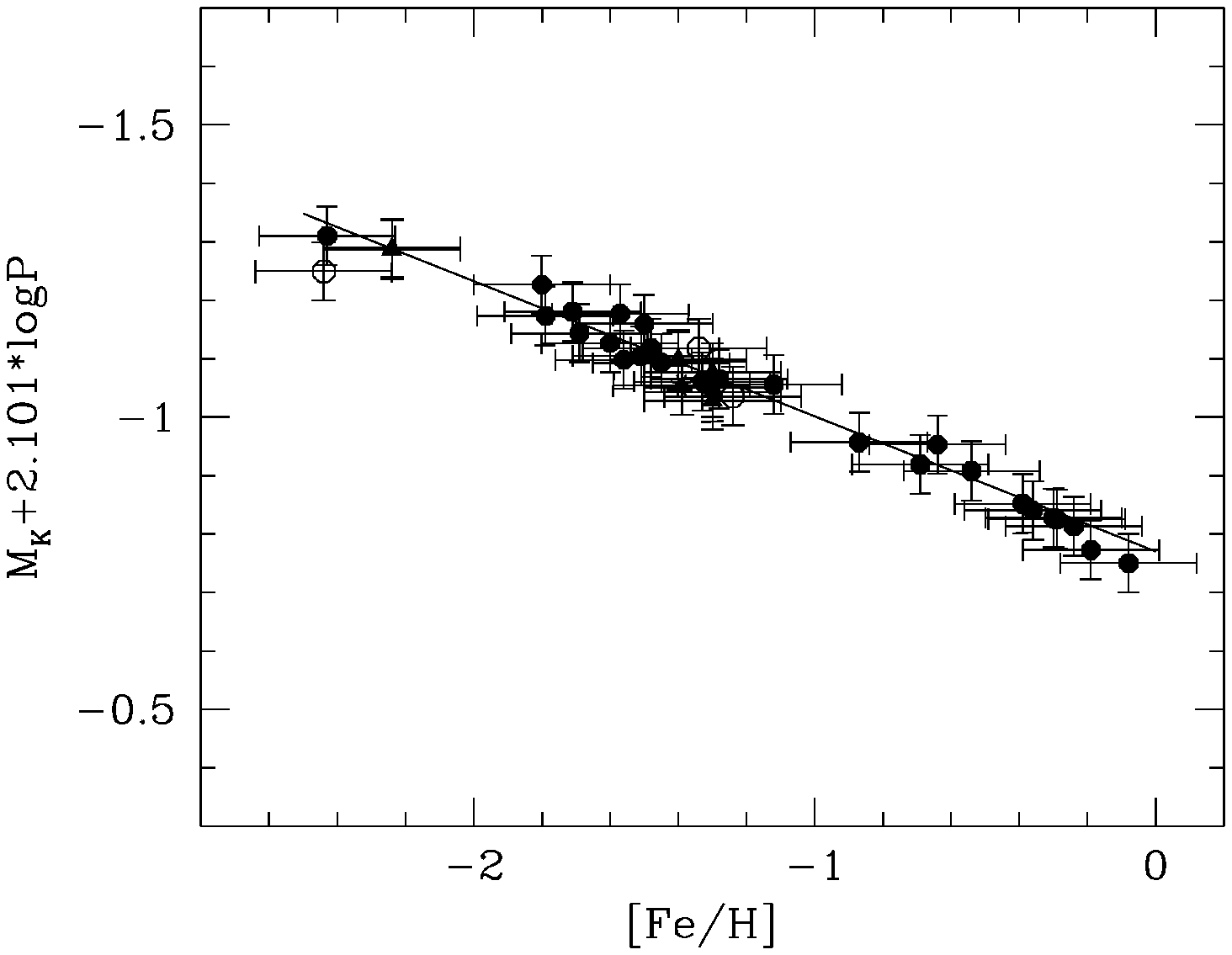}
  \caption{Absolute K magnitudes of field RR Lyrae corrected for the period dependence predicted by pulsation models as a function of metal content. The solid line is the theoretical $P-M_K-[Fe/H]$ relation projected onto a two dimensional plane.}
\end{figure}

\section{Some still open problems in RR Lyrae pulsation models}

Even if the discussion presented in the previous section indicates that 
current nonlinear convective pulsation models are able to nicely reproduce 
many of the observed characteristics of RR Lyrae and to provide relevant 
constraints to important issues of stellar astrophysics, there are a number 
of still debated problems and residual uncertainties.
As stated above, on the basis of the current theoretical scenario we find that 
the $M_V(RR)-[Fe/H]$ relation is not linear over the whole observed 
metallicity range. This prediction is confirmed by evolutionary models (see 
e.g. Cassisi et al. 1998, Demarque et al. 2000, Catelan, Pritzl \& 
Smith 2004) and there are also observational cases that support this 
prediction (see e.g. the results obtained by Rey et al. 2000 for the RR 
Lyrae in Omega Cen). However we find a clear evidence of linearity of the 
 $M_V(RR)-[Fe/H]$ relation for example in the LMC (Clementini et al. 2003). The true 
nature of this correlation and its link with the Oosterhoff dichotomy 
phenomenon definitely deserves further theoretical attention.
Another open problem is the modeling of light curves 
for cool RR Lyrae. In several cases, both in the Galactic field and in GGCs,
 we have found that models are unable to properly reproduce the morphology of 
the light curves close to the red edge of the instability strip.  Given the higher sensitivity to uncertainties affecting the convective transfer, of the pulsation properties of RR Lyrae in the red part of the strip, these results support the idea that further improvements in the treatment of turbulent 
convection are required.

Finally we still lack in the literature a systematic investigation of the 
effect of a possible He enhancement on the properties of RR Lyrae models. 
Very long period RR Lyrae variables have been observed in peculiar GGCs for 
which He enhancement has been invoked (e.g. NGC6441, NGC6388, see Busso et al. 2007; D'Antona \& Ventura 2007).

\begin{figure}
\resizebox{18pc}{!}{\includegraphics{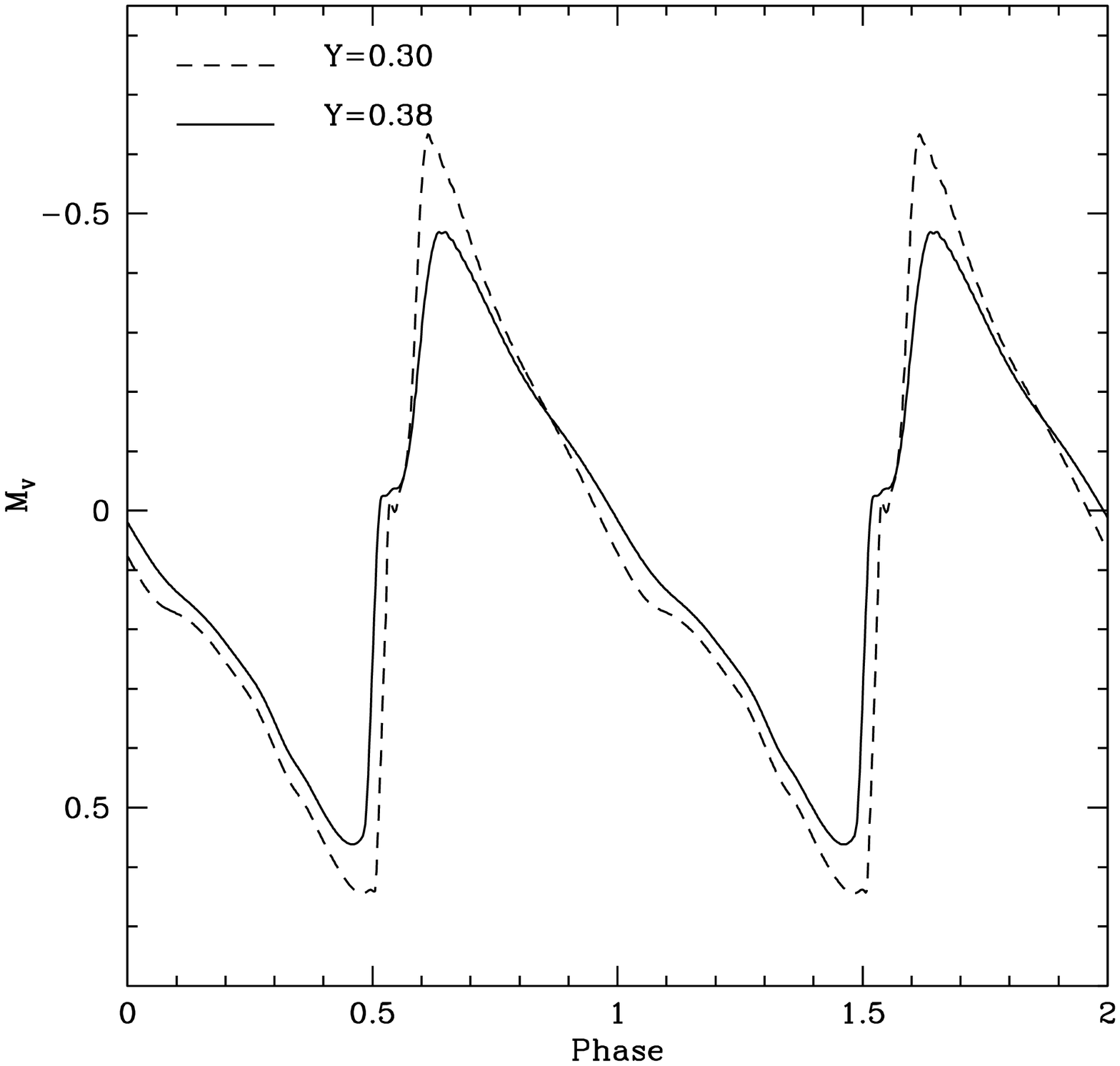}}  
\resizebox{18pc}{!}{\includegraphics{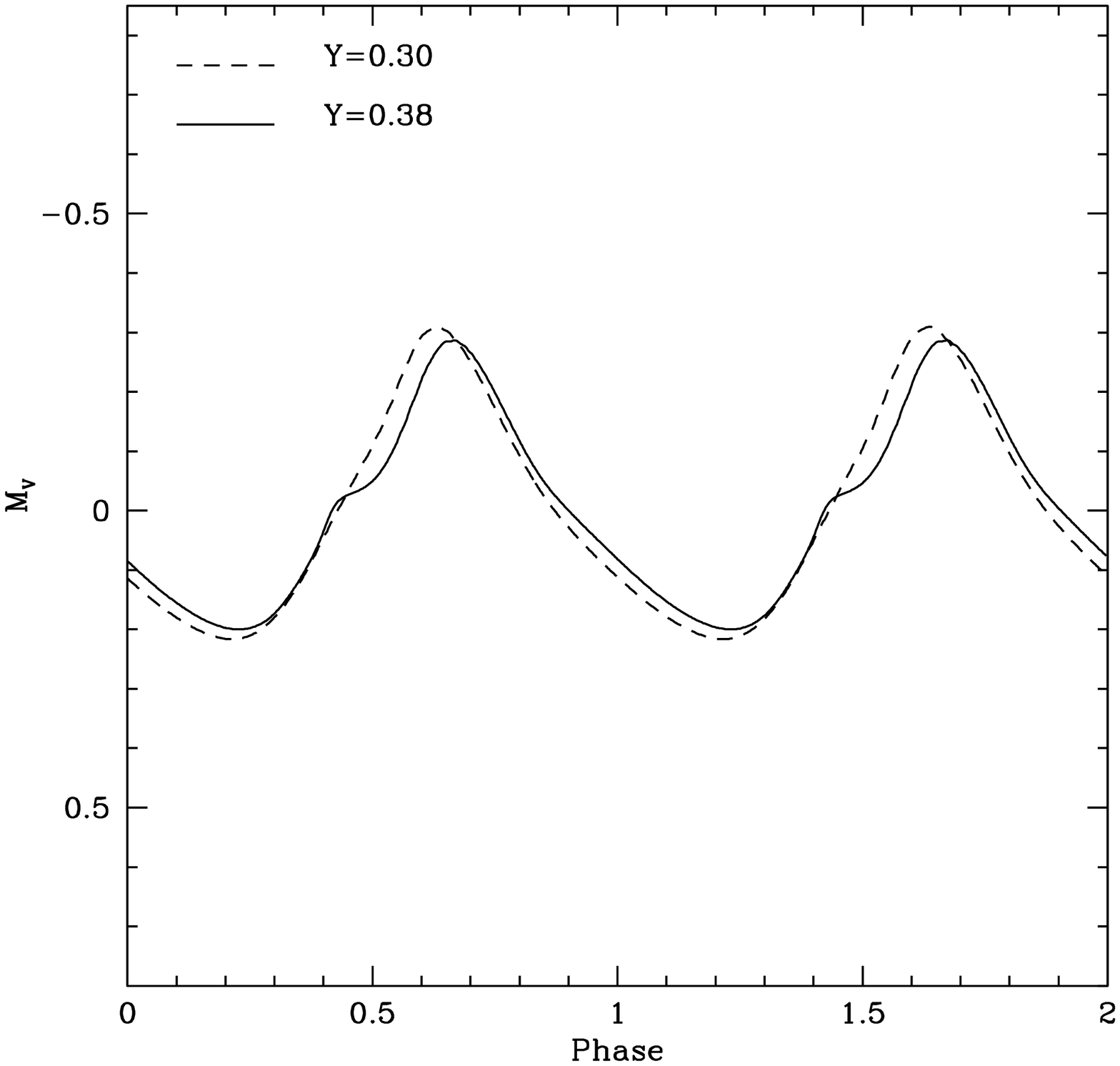}}
\caption{Comparison between the predicted light curves at Y=0.30 and the ones obtained at Y=0.38 for a fundamental (left panel) and first overtone (right panel) model with $Z=0.001$, M=0.60 $M_{\odot}$, $\log{L/L_{\odot}}=1.9$.}
\end{figure}

 \begin{figure}
  \includegraphics[height=.5\textheight]{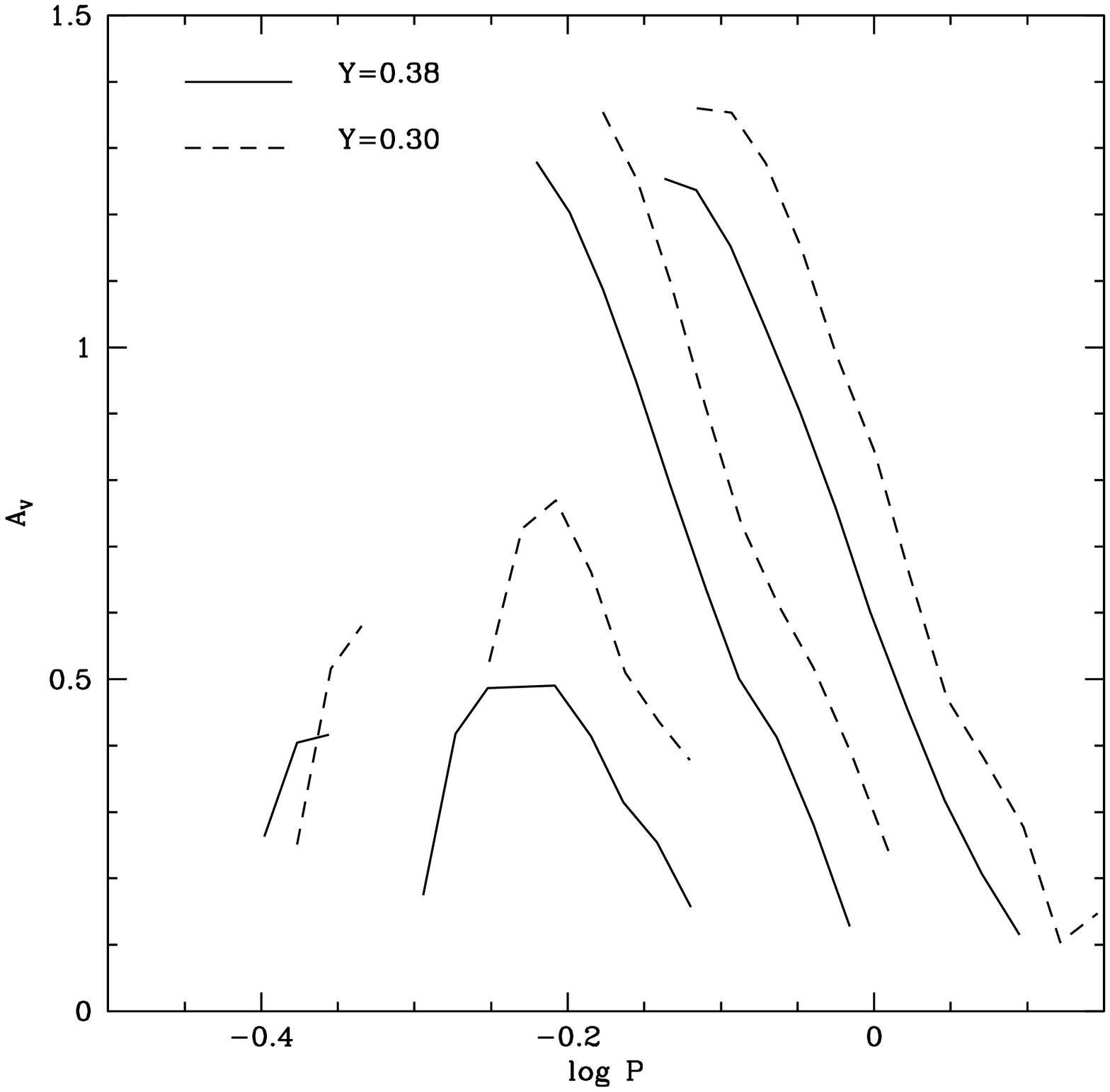}
  \caption{Effect of a variation of Y from 0.30 to 0.38 on the Bailey diagram topology. The other model characteristics are the same as in the previous figure.}
\end{figure}

Model computations for Y higher than 0.30 are in progress but we show in 
Fig. 10  and 11 some preliminary results. From these plots we notice that 
moving from Y=0.30 to Y=0.38 the pulsation amplitudes decrease significantly 
and the light curve morphology can also be modified as exemplified by the 
first overtone case (Fig. 9, right panel). These results imply that the study of the pulsation properties of the RR Lyrae contained in these peculiar clusters might in principle be used to constrain the occurrence of He enhancement.

\section{Conclusions}

RR Lyrae are very important standard candles and tracers of the oldest stellar 
populations. This implies that we need accurate pulsation models to interpret 
their observed characteristics.
Only with nonlinear pulsation models including a non-local time-dependent 
treatment of convection we are able to reproduce all the relevant 
observables of radial pulsation. Important results have been obtained from 
these models and new techniques have been devised to exploit their predictive 
capabilities.
However there a number of physical phenomena that either require the modeling 
of nonradial modes in addition to the radial ones or need the inclusion of 
additional physical inputs.
Moreover, even in the context of the above discussed nonlinear convective 
radial models, there are some still open problems that deserve further 
theoretical investigation. In particular to obtain a successful modeling
of the light and radial velocity curves for RR Lyrae located close to the red 
edge of the instability strip we most likely need a more sophisticated 
treatment of turbulent convection.



\begin{theacknowledgments}
It is a pleasure to thank G. Bono and G. Clementini for their valuable comments.
\end{theacknowledgments}



\bibliographystyle{aipproc}   





\end{document}